\documentstyle[prb,multicol,aps,epsf]{revtex}
\begin{document}
\draft

\title{Thermal Transport as a Probe of Localized Charge and 
Lattice Distortions in Manganites and Cuprates} 
\author{J. L. Cohn} 
\address{Department of Physics , University of Miami, Coral Gables, FL 33124}

\maketitle

\begin{abstract}
The thermal resistivity of manganites and cuprates 
is shown to correlate with local lattice distortions and the
presence of localized holes as determined from neutron diffraction 
and nuclear quadrupole resonance.  Oxygen doping dependent
studies of heat conduction in Y-123 and Hg cuprates
suggest that a fraction of the holes in the CuO$_2$ planes are localized
near the planar hole concentration $p=1/8$.  The results are consistent
with the formation of short-ranged, static stripe domains that are pinned
near oxygen vacancy clusters.

\end{abstract}
\pacs{PACS numbers: 74.72.-h, 74.25.Fy, 74.25.Bt, 74.25.Dw}

\begin{multicols}{2}
\narrowtext

Manganite perovskites and cuprate superconductors 
are both typified by local lattice distortions.
In the colossal magnetoresistance manganites correlations
of the local structure with charge transport 
through the insulator-metal transition demonstrate that 
the distortions are associated with localized holes 
(polarons).\cite{ManganiteDistortions,BoothNew} 
In the cuprates the distortions are smaller in magnitude
and are not as obviously related to the localization of doped holes. 
Nuclear magnetic and quadrupole resonance (NMR and NQR) 
studies\cite{Hammel,HamScal} of La$_{1-x}$Sr$_x$CuO$_4$ (La-214) and 
YBa$_2$Cu$_3$O$_{6+x}$ (Y-123) indicate the presence of localized holes 
in the CuO$_2$ planes.  Whether localized holes are a general feature 
of cuprates, organize in
charge/spin stripes,\cite{CuprateStripes} 
or contribute to the normal-state pseudogap\cite{Loram,PseudoGap} are 
fundamental issues of current interest. 

Here we demonstrate that the thermal resistivity 
of manganites and in the normal state of cuprates is a sensitive 
measure of lattice distortions.  For the cuprates the change in slope of 
the thermal conductivity ($\kappa$) at $T_c$ is also shown to be a measure of the
difference between the normal- 
and superconducting-state low-energy electronic spectral weight
(proportional to the superfluid density).  The phase behavior of
these two thermal transport parameters for YBa$_2$Cu$_3$O$_{6+x}$ 
(Y-123) and Hg cuprates reveal anomalies near 
planar hole concentrations $p=1/8$ which are attributed to localized holes 
in the CuO$_2$ planes.  The implications are discussed.

\begin{figure}
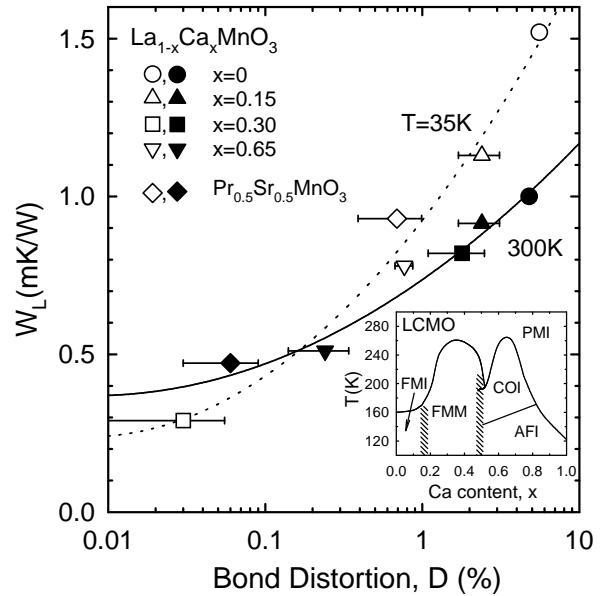

\vskip -.9in
\caption{Lattice thermal resistivity {\it vs} Mn-O bond distortion (see text) for
manganite polycrystals (Ref.~\protect{\onlinecite{CohnManganites}}).  
The phase diagram for La$_{1-x}$Ca$_x$MnO$_3$ is 
shown in the inset, delineating the paramagnetic insulating (PMI), ferromagnetic
insulating (FMI), ferromagnetic metallic (FMM), charge-ordered insulating (COI), and
antiferromagnetic insulating (AFI) regimes.}  
\label{Manganites}
\end{figure}

The thermal resistivity\cite{CohnManganites} of undoped LaMnO$_3$ is 
surprisingly high for a crystalline
insulator ($\sim1$mK/W), approaching the theoretical maximum value\cite{Cahill} 
and indicative of a high degree of disorder.  Such behavior in crystalline materials 
is characteristic of random, noncentral distortions of the lattice, 
attributable in the manganites to Jahn-Teller distortions of the MnO$_6$ octahedra.
Figure~\ref{Manganites} shows that the lattice thermal resistivity,
$W_L=\kappa_L^{-1}$, of doped compounds correlates with a measure of the 
bond disorder determined from neutron diffraction, 
$D\equiv(1/3)\sum_{i=1}^3|(u_i-\bar u)/\bar u|\times 100$, with $u_i$ the 
Mn-O bond lengths, and $\bar u=(u_1u_2u_3)^{1/3}$.   
The correlation is especially convincing given that $D$ is 
dramatically altered by the ferromagnetic and charge-ordering transitions 
upon cooling from the paramagnetic insulating state (Fig.~\ref{Manganites}, inset).
For example, Pr$_{0.5}$Sr$_{0.5}$MnO$_3$ has the smallest $D$ at
300K, but one of the largest at 35K; the reverse is true 
for La$_{0.7}$Ca$_{0.3}$MnO$_3$.  The various phase transitions all involve
modifications of the local structure which correlate with hole itinerancy and the 
magnetism, confirming their polaronic origin.

\begin{figure}
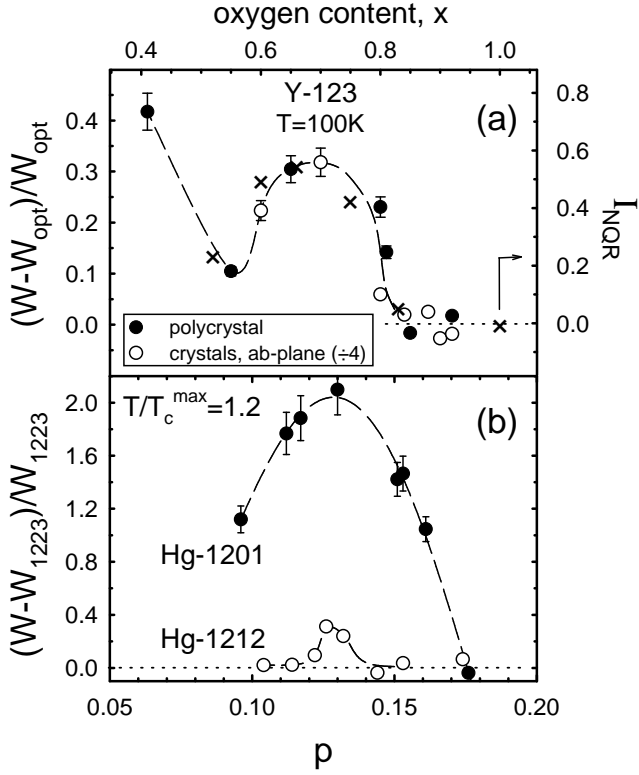

\caption{(a) Thermal resistivity at $T$=100K relative to that at $p_{opt}=0.16$  
for Y-123 polycrystals and the ab-plane of single crystals 
(Ref.~\protect{\onlinecite{Popoviciu}}) [$W_{opt}=0.21$mK/W(0.08mK/W) for the 
polycrystal (crystals)].  Also plotted ($\times$'s) 
is the relative intensity of anomalous $^{63}$Cu NQR signals 
(Ref.~\protect{\onlinecite{HamScal}}).  The dashed curve is a 
guide to the eye. (b) Thermal resistivity for Hg-1201 and Hg-1212 polycrystals 
relative to that of Hg-1223 (Ref.~\protect{\onlinecite{CohnHg}}).
Dashed curves are guides.}  
\label{CuprateW}
\end{figure}

We have conducted extensive studies of $\kappa$ {\it vs} temperature and
oxygen doping in polycrystalline and single-crystal (ab-plane) Y-123 and 
Hg cuprates.\cite{Popoviciu,CohnHg}  Figure~\ref{CuprateW}~(a) shows,  
for Y-123 crystals and polycrystals at $T$=100K, the doping dependence of the 
thermal resistivities, expressed as relative changes from values at optimum doping ($p$=0.16).   
The hole concentration per planar Cu atom, $p$, was determined from thermopower
measurements and the empirical relations with bond va-\break
\vskip .02in
\centerline{\epsfxsize=3.25in,\epsfbox[50 200 520 780]{ericef1.psc}}
\centerline{\epsfxsize=3.2in\epsfbox[50 60 540 770]{ericef2.psc}}
\vskip -.7in
\noindent
lence sums established by 
Tallon {\it et al.}\cite{PfromTEP}  In spite of additional thermal resistance 
due to $c$-axis heat flow and granularity in the polycrystal, the doping dependence 
is essentially the same as that of the ab-plane of crystals.
The behavior thus reflects the intrinsic oxygen doping dependence of the thermal resistivity 
for heat flow in the planes.
\par
Given that the electronic thermal conductivity ($\kappa_e$) at $T>T_c$ 
in optimally doped, single-crystal Y-123
represents only $\sim$10\% of the total 
$\kappa=\kappa_e+\kappa_L$,\cite{Krishana} the data 
imply a lattice thermal resistivity that is anomalously
enhanced about $p=1/8$.  
This enhancement correlates quite well with the relative intensity 
[$I_{NQR}$, Fig.~\ref{CuprateW}~(a)] of anomalous $^{63}$Cu NQR signals 
from planar Cu (below 30.8MHz), attributed to localized holes in 
the planes.\cite{HamScal}  

Similarly enhanced thermal resistivities about $p$=1/8 are observed in 
studies\cite{CohnHg} of HgBa$_2$Ca$_{m-1}$Cu$_m$O$_{2m+2+\delta}$ 
[Hg-1201 ($m$=1), Hg-1212 ($m$=2), Hg-1223 ($m$=3)].  As we discuss further below,
Hg-1223 shows no apparent anomaly, so we use its $W(p)$ curve as a reference
to plot the relative thermal resistivities of the other compounds in Fig.~\ref{CuprateW}~(b).  
These data provide some insight into the role of oxygen vacancies in this phenomenon.
A single HgO$_{\delta}$ layer
per unit cell contributes charge to $m$ planes in 
Hg-12($m$-1)$m$ so that the oxygen occupancy $\delta$ increases with $m$.
At optimum doping,\cite{Occupancy} $\delta_{opt}\simeq0.18$ (Hg-1201), 
0.35 (Hg-1212) and 0.41 (Hg-1223). 
The anomalous thermal resistivity and the range in $p$ over which it occurs
increase with the vacancy concentration, $1-\delta$.

\begin{figure}
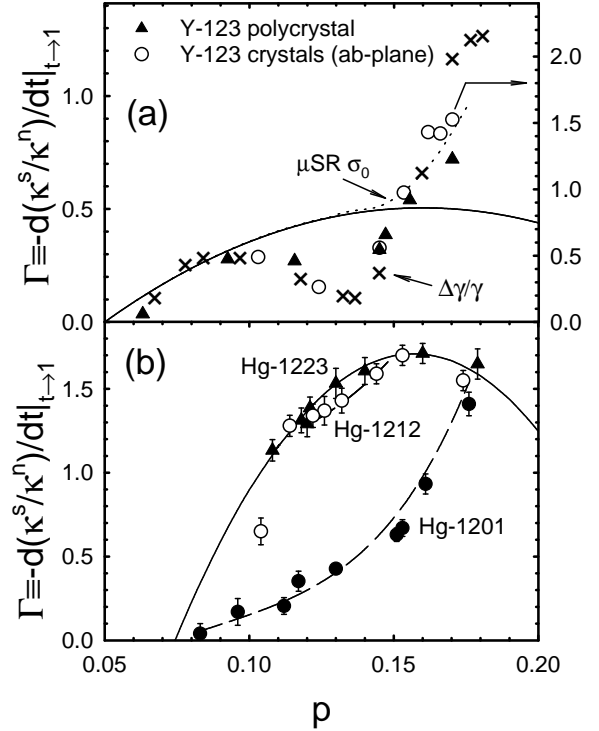

\caption{(a) The normalized slope 
change in $\kappa(T)$ at $T_c$ {\it vs} doping for each of the 
Y-123 specimens from Fig.~\protect{\ref{CuprateW}}~(a).  
Also shown are the normalized electronic specific heat jump, 
$\Delta\gamma/\gamma$,
from Loram {\it et al.} (Ref.~\protect{\onlinecite{Loram}}) ($\times$'s), and the 
$\mu$SR depolarization rate 
from Ref.~\protect{\onlinecite{muSRY123}} (dotted curve), 
divided by 1.4 and 2.7, respectively,
and referred to the right ordinate.  The solid line is 
$0.51-41.7(p-0.16)^2$. (b) The normalized slope change in $\kappa(T)$ 
at $T_c$ {\it vs} doping for Hg
cuprates.  The solid line is $1.71-250(p-0.157)^2$.  Dashed curves are guides.}  
\label{GammaofP}
\end{figure}

A further indication that the lattice distortions responsible for 
enhanced thermal resistivity
are associated with localized holes
comes from the behavior of the slope change in $\kappa$ at $T_c$.
Figure~\ref{GammaofP} shows
the phase behavior of the dimensionless superconducting-state slope change\cite{Popoviciu} 
defined as $\Gamma\equiv -d(\kappa^s/\kappa^n)/dt|_{t=1}$, where $t=T/T_c$ and 
$\kappa^s (\kappa^n)$ is the thermal conductivity in 
the superconducting (normal) state. $\Gamma$ measures the change 
in scattering of heat carriers (electrons and phonons) 
induced by superconductivity.  Particularly striking 
is the similarity between the electronic 
specific heat jump for Y-123,\cite{Loram} 
$\Delta\gamma/\gamma(p)$, 
and $\Gamma(p)$; both quantities
exhibit minima near $p=1/8$. 

Though the origin of the slope change in 
$\kappa$ at $T_c$ (i.e. electronic or phononic) 
has been a subject of some debate, the results of
Fig.~\ref{GammaofP}~(a) demonstrate unambiguously that $\Gamma$,
like $\Delta\gamma/\gamma$, measures
the difference between\break
\vskip .15in
\centerline{\epsfxsize=3.25in\epsfbox[50 60 540 770]{ericef3.psc}}
\vskip -1in
\noindent 
the normal- 
and superconducting-state low-energy electronic spectral weight.  
The muon spin rotation ($\mu$SR) depolarization rate 
($\sigma_0$),\cite{muSRY123} 
a measure of the superfluid density, matches the data for $\Gamma$
and $\Delta\gamma/\gamma$ at $p<0.09$ and $p>0.15$ when suitably scaled
[dotted curve in Fig.~\ref{GammaofP}~(a)].   The sharp rise in $\sigma_0$
for $p>0.15$ is due to the superconducting condensate on oxygen-filled chains,\cite{muSRY123}
motivating the solid curve in Fig.~\ref{GammaofP} (a)] as an estimate of
$\Gamma(p)$ for the CuO$_2$ planes alone in Y-123.  The range of $p$ over
which $\Gamma$ and $\Delta\gamma/\gamma$ deviate from the $\mu$SR curve {\it coincides} with the
range of enhanced thermal resistivity.  The suppressed
transfer of spectral weight below $T_c$ is
also consistent with the localization of a fraction of planar holes.
That $\sigma_0$ is not also suppressed near $p$=1/8
indicates that hole-localized domains do not 
inhibit the formation of a flux lattice in adjacent regions where holes are itinerant.

The $\Gamma(p)$ data for the Hg cuprates [Fig.~\ref{GammaofP}~(b)] 
tell a similar story.
For Hg-1223 $\Gamma(p)$ describes an inverted parabola centered near 
$p=0.16$ (solid curve),
with the data for Hg-1212 and Hg-1201 suppressed from this curve
in ranges of $p$ that coincide with those where 
$W(p)$ is enhanced.  Evidently the concentration of localized holes in Hg-1223
is sufficiently small that neither $\Gamma$ nor $W$ exhibit anomalies.

If our interpretation about localized holes is correct, the Hg data make it clear 
that oxygen vacancies play only a supporting role. 
The importance of 1/8th doping implies that the phenomenon
involves an excitation of the CuO$_2$ planes that is commensurate with 
the lattice.  A plausible candidate is a  
small domain of static stripe order,\cite{CuprateStripes} nucleated via pinning by a 
vacancy-induced mechanism.  Recent Raman scattering studies of 
optimally-doped Hg compounds\cite{HgRaman} implicate
vacancy clusters in pinning: the oscillator strength of the 
590 cm$^{-1}$ mode, 
attributed to c-axis vibrations of apical oxygen in an 
environment of four vacant nearest-neighbor dopant sites,
scales with the magnitudes of the $W$ enhancement and $\Gamma$ 
suppression.  Oxygen vacancies in both Y-123 and Hg cuprates\cite{Occupancy,Jorgensen} 
displace neighboring Ba$^{2+}$ toward the 
CuO$_2$ planes.  The bonding of Ba with apical oxygen could lead to local distortions
of the CuO polyhedra that pin a charge stripe.  Alternatively, a magnetic mechanism is 
possible.  In the Hg cuprates the shift of Ba atoms associated with a cluster 
of four vacancies nearest a CuO polyhedron will induce a positive potential in the planes 
that inhibits its occupation by a hole, thereby suppressing spin fluctuations and
possibly fixing a Cu$^{2+}$ spin at the site.  
Larger vacancy clusters, surrounding adjacent CuO
polyhedra oriented along $\langle 1 0 0\rangle$ directions, may   
induce several spins to order antiferromagnetically via superexchange.
The presence of this spin-chain fragment would favor 
charge/spin segregation that is characteristic of stripe order. 

Consider the relevant length scales.
If randomly distributed, static stripe domains are to scatter 
phonons, their in-plane extent must be less 
than the phonon mean-free-path, $\Lambda$, and their mean separation 
comparable to $\Lambda$.
Expressing $\Lambda^{-1}$ as a sum
of terms for scattering by these domains and by all other processes, 
$\Lambda^{-1}=\Lambda^{-1}_{str}+\Lambda^{-1}_{other}$, we use the in-plane
thermal resistivity for Y-123 and kinetic theory to find\cite{NoteonMFP}
$\Lambda_{str}\simeq [3/C_Lv\Delta W(p$=$1/8)]\simeq 
70{\rm\AA}$ as an estimate of the separation between domains at $p$=1/8.
We can make a rough estimate of the fractional area of the planes 
having static stripes by taking the typical domain size to be 
$2a\times8a$ ($a$ is the lattice constant), the stripe unit cell  
suggested\cite{CuprateStripes} for (La,Nd)-214.  This yields a fraction
$16(a/\Lambda_{str})^2\simeq 0.05$.  Our data suggest that 
this fraction is substantially higher in Hg-1201, but apparently below the 
2-D percolation threshold of 0.50.  This presumably explains 
why no substantial $T_c$ suppression is observed near $p$=1/8 in 
Y-123 or the Hg cuprates, in contrast to the case of
(La,Nd)-124; stripe domains in the latter system have longer-range order
($\geq 170\rm{\AA}$).\cite{CuprateStripes}  

This work was supported by NSF Grant No.~DMR-9631236.

\end{multicols}
\end{document}